\documentclass[12pt]{emulateapj}

\usepackage{graphics}

\renewcommand{\min}{\mbox{$^m$}}
\renewcommand{\sec}{\mbox{$^s$}}
\renewcommand{\deg}{\mbox{$^{\circ}$}}

\def\deg{\ifmmode^\circ\else$^\circ$\fi}    %Degree sign%

\def\hper{\ifmmode \rlap.^{h}\else $\rlap{.}^h$\fi} 
\def\sper{\ifmmode \rlap.^{s}\else $\rlap{.}^s$\fi}    %Superscript 's' over per
\def\Sec{${}^{\prime\prime}$\llap{.}}
\def\deg{${}^\circ$}
\def\min{${}^{\prime}$}
\def\sec{${}^{\prime\prime}$}

\def\abun#1{\hbox{\rm [#1]}}
\def\ngc#1{NGC$\,$#1}

\def\today{\number\year\space \ifcase\month\or
  January\or February\or March\or April\or May\or June\or
  July\or August\or September\or October\or November\or December\fi
  \space\number\day}
\def\now{\number\year\space \ifcase\month\or
  January\or February\or March\or April\or May\or June\or
  July\or August\or September\or October\or November\or December\fi
  \space\number\day .\number\time}

\shorttitle{On the radial distribution of HB stars in \ngc{2808}\altaffilmark{1}} 
\shortauthors{Iannicola et al.}

\begin{document}
\title{On the radial distribution of horizontal branch stars in  \ngc{2808}\altaffilmark{1}}

\author{
G.~Iannicola\altaffilmark{2},
M.~Monelli\altaffilmark{3},
G.~Bono\altaffilmark{2,4},
P.B.~Stetson\altaffilmark{5}
R.~Buonanno\altaffilmark{4},
A.~Calamida\altaffilmark{6},
M.~Zoccali\altaffilmark{7}, 
F.~Caputo\altaffilmark{2}, 
M.~Castellani\altaffilmark{2}, 
C.E.~Corsi\altaffilmark{2},
M.~Dall'Ora\altaffilmark{8}, 
A.~Di Cecco\altaffilmark{2,4}, 
S.~Degl'Innocenti\altaffilmark{9,10}, 
I.~Ferraro\altaffilmark{2},
M.~Nonino\altaffilmark{11}, 
A.~Pietrinferni\altaffilmark{12}, 
L.~Pulone\altaffilmark{2}, 
P.G.~Prada Moroni\altaffilmark{9,10},
M.~Romaniello\altaffilmark{6}, 
N.~Sanna\altaffilmark{2,4}, and 
A.R.~Walker\altaffilmark{13} 
}

\altaffiltext{1}
{Based on observations collected with the ACS, WFPC2 and STIS 
on board of the Hubble Space Telescope.}
  \altaffiltext{2}{INAF--OAR, via Frascati 33, Monte Porzio Catone, Rome, Italy; giacinto@mporzio.astro.it} 
  \altaffiltext{3}{IAC, Calle Via Lactea, E38200 La Laguna, Tenerife, Spain}
  \altaffiltext{4}{UniToV, Dipartimento di Fisica, via della Ricerca Scientifica 1, 00133 Rome, Italy} 
  \altaffiltext{5}{DAO--HIA, NRC, 5071 West Saanich Road, Victoria, BC V9E 2E7, Canada} 
  \altaffiltext{6}{ESO, Karl-Schwarzschild-Str. 2, 85748 Garching bei Munchen, Germany}
  \altaffiltext{7}{PUC, Departamento de Astronomía y Astrofísica, Casilla 306, Santiago 22, Chile} 
  \altaffiltext{8}{INAF--OACN, via Moiariello 16, 80131 Napoli, Italy} 
  \altaffiltext{9}{Univ. Pisa,  Dipartimento di Fisica, Largo B. Pontecorvo 2, 56127 Pisa, Italy}
  \altaffiltext{10}{INFN--Pisa, via E. Fermi 2, 56127 Pisa, Italy}
  \altaffiltext{11}{INAF--OAT, via G.B. Tiepolo 11, 40131 Trieste, Italy} 
  \altaffiltext{12}{INAF--OACTe, via M. Maggini, 64100 Teramo, Italy}
  \altaffiltext{13}{CTIO--NOAO, Casilla 603, La Serena, Chile} 

\date{\centering drafted \today\ / Received / Accepted }

\begin{abstract}
We present accurate new ultraviolet and optical {\it BVI\/} photometry for the
Galactic globular cluster \ngc{2808}, based on both ground-based and archival
HST imagery.  From this we have selected a sample of $\sim$2,000 HB stars; given
the extensive wavelength range considered and the combination of both
high-angular-resolution and wide-field photometric coverage, our sample should
be minimally biased.  We divide the HB stars into three radial bins and find
that the relative fractions of cool, hot and extreme HB stars do not change
radically when moving from the center to the outskirts of the cluster: 
the difference is typically smaller than $\sim$2$\sigma$.  
These results argue against the presence of {\it strong\/} radial 
differentiation among any stellar subpopulations having distinctly 
different helium abundances.  
% Point A 
The ratio between HB and RG stars brighter than the ZAHB steadly increases
when moving from the innermost to the outermost cluster regions. The difference
is larger than $\sim$4$\sigma$ and indicates a deficiency of bright RGs in the
outskirts of the cluster. 
\end{abstract}

\maketitle

%%%%%%%%%%%%%%%%%%%%%%%%%%%%%%%%%%%%%%%%%%%%%%%%%%%%%%%%%%%%%%%%%%%%%
\section{Introduction}

The Galactic globular cluster (GGC) \ngc{2808} is a very interesting stellar
system. Space- and ground-based optical photometry show that both the red giant
branch (RGB) and the main sequence (MS) show spreads in color larger than can be
explained  by intrinsic photometric errors (Bedin et al.\ 2000).  However,
recent accurate high-resolution spectroscopic measurements of $\sim$120 RGs
(Carretta 2006) indicate that no spread in metal abundance is present
($\abun{Fe/H}=-1.10\pm 0.065$). The observational scenario was enriched by the
detection of a fringe of stars blueward of the canonical MS by D'Antona et al.\
(2005). This evidence was solidified by the identification of a triple
MS by Piotto et al.\ (2007) using accurate photometry of an off-center
field collected with the Advanced Camera for Surveys (ACS) on board the Hubble
Space Telescope (HST). The authors proposed that this cluster had experienced
different episodes of star formation with significant helium enrichment ($0.30
\le Y \le 0.40$).     

\ngc{2808} also shows an extended blue HB with multiple distinct components,
and a helium enrichment sufficient to explain the presence of multiple MSs 
had already been proposed to explain this morphology (D'Antona et al.\
2002,2005; Lee et al.\ 2005). More recently, D'Ercole et al.\ (2008) have
performed detailed simulations of the formation and the dynamical evolution of
GCs hosting multiple stellar populations. They found that a substantial fraction
of first generation stars are lost during the early cluster evolution. Moreover,
the resulting radial profile of the ratio between second and first generation MS
stars is characterized by a flat trend in the inner regions with a decrease
(i.e., the first generation relatively more important compared to the second) in
the outer cluster regions (see especially their Fig.~18).  

From analysis of optical and UV Wide Field Planetary Camera 2 (WFPC2) images, 
Castellani et al.\ (2006) found  
that the ratio between HB and RG stars brighter than the HB luminosity level 
(the classical $R$ parameter) in \ngc{2808} increases when
moving from the core to the outermost regions.  A deficiency of bright RGs in
this cluster was suggested by Sandquist \& Martel (2007) on the basis of HST 
images.  
% Point B 
Peculiar radial distributions of HB stars in $\omega$~Centauri have been suggested
by Castellani et al.\ (2007).
More recently, Zoccali et al.\ (2009) argued that the fainter (peculiar) of 
the two subgiant branches detected in \ngc{1851} (Milone et al.\ 2008; 
Cassisi et al.\ 2008) disappears at radial distances larger than 2.4 arcmin.     
 
In this investigation we focus our attention on the luminosity function  
of HB stars in \ngc{2808}.

%-------------------------------------------------------------------------
\section{Observations and data reduction}

In order to address these issues we took advantage of several 
UV and optical datasets collected with both HST and ground-based 
telescopes. These include UV data from the Space Telescope Imaging 
Spectrograph (STIS, 8511, PI: P. Goudfrooij) in far-UV 
(FUV, $F25QTZ$, $\lambda_c=1590$, FWHM=220 \AA) and near-UV 
(NUV, $F25CN270$, $\lambda_c=2700$, FWHM=350 \AA) bandpasses 
(Brown et al.\ 2001; Dieball et al.\ 2005,2009). Individual images cover a 
field of view (FoV) of 25\sec$\times$25\sec, and the pixel scale is 
0\Sec0248/px. The entire data set consists of 18 FUV and 18 NUV 
images with exposure times ranging from 480 to 538\,s located across 
the cluster center (pointing $\alpha$, see Fig~1). 
We also used data from the ACS High Resolution Channel 
(ACS HRC, 10335, PI: H. Ford) in $F435W$ 
($24\times 135\,$s) and $F555W$ ($4\times 50\,$s). 
Individual images cover a FoV of 29\sec$\times$26\sec and 
the pixel size is 0\Sec028$\times$0\Sec025. These images, too,
are located across the cluster center (pointing $\beta$, 
Sandquist \& Martel 2007). We also used data collected with
ACS Wide Field Camera (ACS WFC, 10775, PI: A. Sarajedini) in 
$F606W$ ($5\times360\,$s and $1\times 23\,$s) and $F814W$ 
($5\times370\,$s + $1\times23\,$s). Individual images span
202\sec$\times$202\sec at 0\Sec05/px; these are slightly dithered and are also
located across the cluster center (pointing $\epsilon$). These data were
supplemented with optical/UV data images from the WFPC2: 
pointing $\gamma$, located near the cluster center--- $F218W$
($1\times 1600\,$s + $1\times 1700\,$s), $F439W$ ($2\times 230\,$s +
$1\times 50\,$s), $F555W$ ($1\times 7\,$ + $1\times 50\,$s) from proposal 6095
(PI: S.G. Djorgovski; also Bedin et al.\ 2000; Castellani et al.\ 2006;
Sandquist \& Hess 2008);  pointing $\delta_i$, also close to the cluster center
and partially overlapping $\gamma$--- $F336W$ ($2\times3600\,$s), $F555W$
($3\times 100\,$s + $2\times7\,$s), $F814W$ ($3\times 120\,$s + $2\times3\,$s)
from proposal 6804 (PI: F. Fusi Pecci); pointing $\delta_o$, a couple
of arcminutes southwest of the cluster center---
$F336W$ ($4\times1600\,$s), $F555W$ ($4\times900\,$s + 
$3\times 100\,$s + $2\times7\,$s), $F814W$ ($4\times700\,$s + 
$3\times 120\,$s + $2\times 3\,$s), also from proposal 6804.

For wider coverage of the cluster regions, we collected multiband ({\it UBVI\/})
ground-based archival data covering $\approx$ 20\min$\times$ 15\min around the
center of the cluster (pointing $\zeta$, see Fig.~1). 
%Point C 
The dark area around the cluster centre marks the high density cluster regions. 
A total of 573 CCD images collected
between 1987 January and 2002 February were reduced, but not all of these could
be calibrated because of an insufficiency of standard-star observations. 
Moreover, since some of these data were acquired with mosaic cameras, and since
even for single-chip cameras not all telescope pointings were the same, no
individual star could be measured in all images.  To summarize, calibrated
photometry for any given star is available from at most 18 CCD images in $U$, 26
in $B$, 66 in $V$, and 59 in $I$.  

Photometric reduction of the STIS images was done with {\tt ROMAFOT}, while
photometry of pointing $\gamma$ was obtained with both {\tt ROMAFOT} and {\tt
DAOPHOT/ALLFRAME}. The photometry for all the other data was carried out with
{\tt DAOPHOT/ALLFRAME}.  The final catalog includes $\sim 379,000$ stars with at
least one measurement in two or more different optical bands, $\sim 100,000$ 
stars with at least one $U$ or $F336W$ measurement, and $\sim 4,900$ stars with
at least one shorter-wavelength UV measurement ($FUV$, $NUV$, $F218W$).  

\begin{figure}[!ht]
\begin{center}
\label{fig1}
\includegraphics[height=0.375\textheight,width=0.4\textwidth]{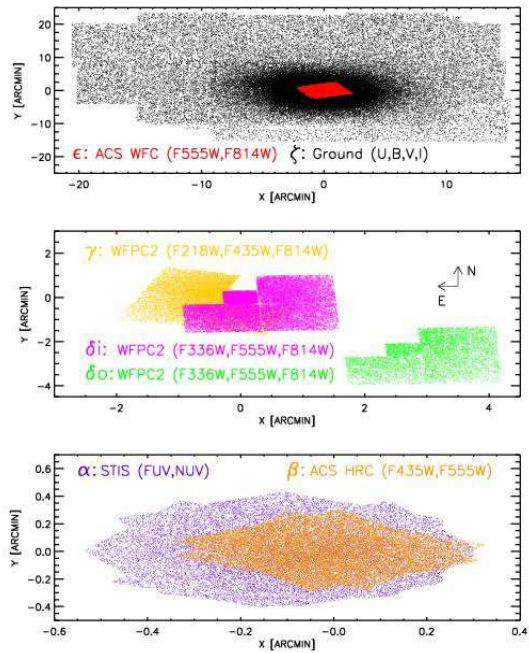}
\caption{
Top -- multiband data collected with ACS@HST (WFC,
$F555W$, $F814W$) and ground-based telescopes ($U,B,V,I$).
Middle -- same as the top, but for data collected with WFPC2@HST
($F218W$, $F336W$, $F555W$, $F814W$).
Bottom -- same as the top, but for data collected with STIS@HST
($FUV$, $NUV$) and with ACS@HST (HRC, $F435W$, $F555W$).
The arrows mark the Northern-Eastern directions.
}
\end{center}
\end{figure} 
The WFPC2 photometry was kept in the Vega system following the prescriptions
suggested by Holtzman et al.\ (1995). The STIS photometry was referred to the
Vega system following the prescriptions by Brown et al.\ (2001) and Dieball et
al.\ (2005).  For homogeneity the ACS $F435W$, $F555W$, $F606W$ and $F814W$
magnitudes were transformed into WFPC2 $F435W$, $F555W$ and $F814W$ magnitudes
on the basis of common stars.  Finally, for the same reason, the ground-based
{\it UBVI\/} magnitudes were transformed into the WFPC2 $F336W$, $F435W$,
$F555W$ and $F814W$ systems. The typical accuracy of the calibration is of the
order of 0.02 mag for all the quoted bands. Here, however, we would like to 
stress that the results of the present paper depend upon {\it star counts\/} 
in different, easily recognized zones of the Color-Magnitude Diagrams (CMDs).  
Accuracy of the absolute calibrations is not a serious consideration for 
the present analysis.

% ******************************************************************************

\section{Results and discussion}

To identify HB stars near the center of the cluster we adopted the optical and
UV results from HST. The reason is twofold.  {\em i)} Detectors with a high
angular resolution are mandatory for accurate photometry in crowded cluster
regions.  Note that the core radius and the half mass radius of \ngc{2808} 
are $r_c=0.26$ and $r_h=0.76$ arcmin 
(Harris 1996\footnote{http://physun.physics.mcmaster.ca/~harris/mwgc.dat}), 
respectively.  Fortunately,
these regions are covered by data sets collected with four different detectors
(STIS, ACS HRC, ACS WFC, WFPC2).   {\em ii)} The optical-UV colors are highly
sensitive to effective temperature, providing the opportunity to properly select
hot HB stars.  For these reasons we gave the highest priority to CMDs based on
optical-UV magnitudes. In particular, we adopted $F555W$,$FUV$-$F555W$;
$F555W$,$NUV$-$F555W$; $F555W$,$F218W$-$F555W$; and $F555W$, $F336W$-$F555W$
CMDs\footnote{Note that we used weighted averages for the magnitudes of stars
present in multiple data sets.}. Data plotted in Fig.~2 show that the difference
in color between  Extreme HB (EHB) and RG stars ranges from $\sim$ 6 (panel d)
to $\sim$ 14 mag (panel a). Moreover, they also show that in the UV bands
(panels a,b,c) we detected not only HB stars but also Blue Stragglers
($17.5\lesssim$$F555W$$\lesssim19$, $1\lesssim$$F218W$-$F555W$$\lesssim2$),
turnoff stars (TO, see the arrow in Fig.~2), main-sequence stars (MS), and a
handful of bright RG stars. To our knowledge these are the deepest optical-UV
CMDs ever collected for a GC.   

\begin{figure*}[!ht]
\begin{center}
\label{fig2}
%\vspace*{-0.45truecm}
\includegraphics[height=0.60\textheight,width=0.35\textwidth,angle=90]{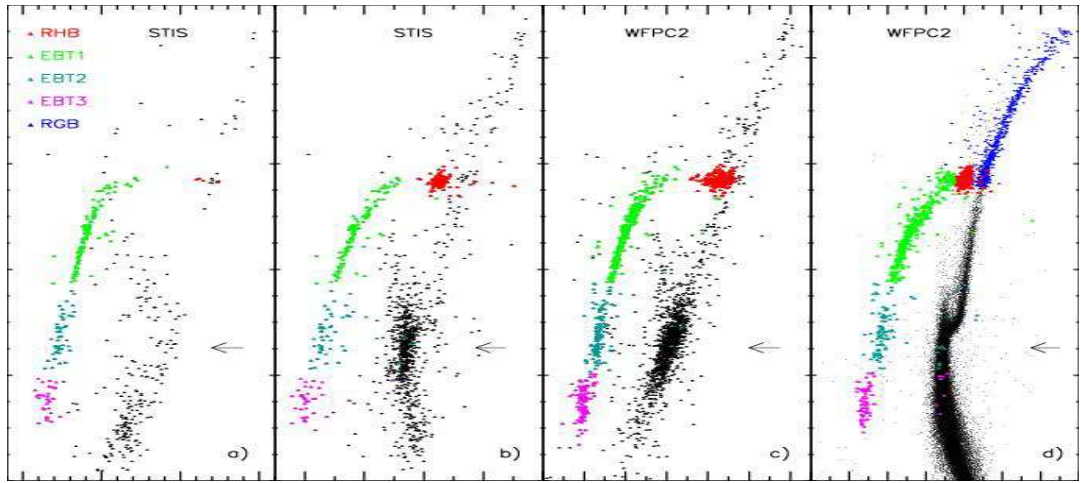}
\vspace*{0.40truecm}
\caption{Optical-UV CMDs of \ngc{2808} based on data collected with HST: 
$F555W$, $FUV$-$F555W$ (a), $F555W$,$NUV$-$F555W$ (b), 
$F555W$,$F218W$-$F555W$ (c), $F555W$,$F336W$-$F555W$ (d).  
Red, green, cyano and purple dots show different HB subgroups, 
while blue dots display RGB stars.
The horizontal arrows mark the luminosity of Tur-Off stars 
($F555W=19.82\pm0.02$ mag).   
}
\end{center}
\end{figure*}

\begin{figure*}[!ht]
\begin{center}
\label{fig3}
\includegraphics[height=0.6\textheight,width=0.35\textwidth,angle=90]{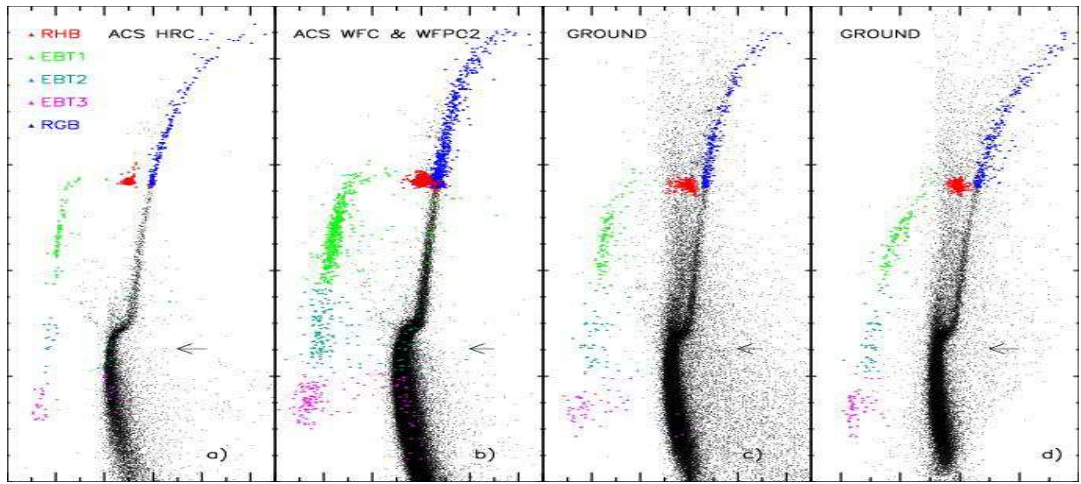}
\vspace*{0.40truecm}
\caption{Same as Fig.~2, but for data sets collected with HST and with 
ground-based telescopes: 
$F555W$, $F439W$-$F555W$ (a), $F555W$,$F555W$-$F814W$ (b), 
$F555W$,$F555W$-$F814W$ (ground, c), $F555W$,$F336W$-$F814W$ (ground, d). 
Red, green, cyano and purple dots mark different subgroups of HB stars. 
The color coding of HB and RGB stars is the same as in Fig.~2. 
Data covering the external cluster regions show the contamination of 
field stars for $14.5 \lesssim$$F555W$$\lesssim 19$, 
$0.4 \lesssim$$F555W$-$F814W$$\lesssim 1.0$ (c) and for 
$14.5 \lesssim$$F555W$$\lesssim 19$, $1 \lesssim$$F336W$-$F814W$$\lesssim 2.0$, 
(d).  
}
\end{center}
\end{figure*}

Following Castellani et al.\ (2006) we define red HB (RHB) stars as those with
colors redder than the RR Lyrae instability strip and with $15.70 \le $F555W$
\le 16.61$ mag.  The HB stars bluer than the instability strip range from 15.8
to 21.6 in $V$ or $F555W$ magnitude, and we confirm the finding of Bedin et al.\ (2000) 
that they can---with minimal ambiguity---be divided into three subsets by
cuts near $V \sim F555W \sim 18.26$ and 19.95.  From brightest to faintest, we
refer to these three subgroups as EBT1, EBT2, and EBT3.  The stars belonging to
EBT3 have been called by different names in the literature (Moehler et al.\
2004; Dalessandro et al.\ 2008), but we adopt our present nomenclature because
we do not want to address their physical nature in this investigation.  Note
that in panel d) the RR Lyrae gap is not clearly identified, so to
distinguish RHB from  EBT1 stars we adopted either the $F435W$-$F555W$ or the
$F555W$-$F814W$ color.     

To avoid possible biases in the selection criteria and to improve the coverage
of the inner cluster regions we also considered purely optical CMDs based on HST
data: in particular, $F555W$,$F439W$-$F555W$ and $F555W$,$F555W$-$F814W$. 
Finally, to cover the area between the inner regions covered by HST data and the
tidal radius ($r_t=15.5$ arcmin) we adopted the
$F555W$,$F555W$-$F814W$ and the $F555W$,$F336W$-$F814W$ CMDs inferred from
ground-based photometry (see panel c,d of Fig.~3).  Data plotted in Fig.~3 show
that the current optical photometry provides very good sampling from the tip of
the RGB down to a couple of magnitudes fainter than the TO.  Despite
\ngc{2808}'s low Galactic latitude ($l^{\hbox{\footnotesize II}} = 282$\deg,
$b^{\hbox{\footnotesize II}} = -11$\deg, the EBT1, EBT2, and EBT3 samples should
be minimally contaminated by field stars, since stars this hot and faint are
rare); conversely, RHB stars have $F555W$-$F814W$ and $F336W$-$F814W$ colors
similar to common field stars.  

\begin{figure}[!ht]
\begin{center}
\label{fig4}
\includegraphics[height=0.4\textheight,width=0.40\textwidth]{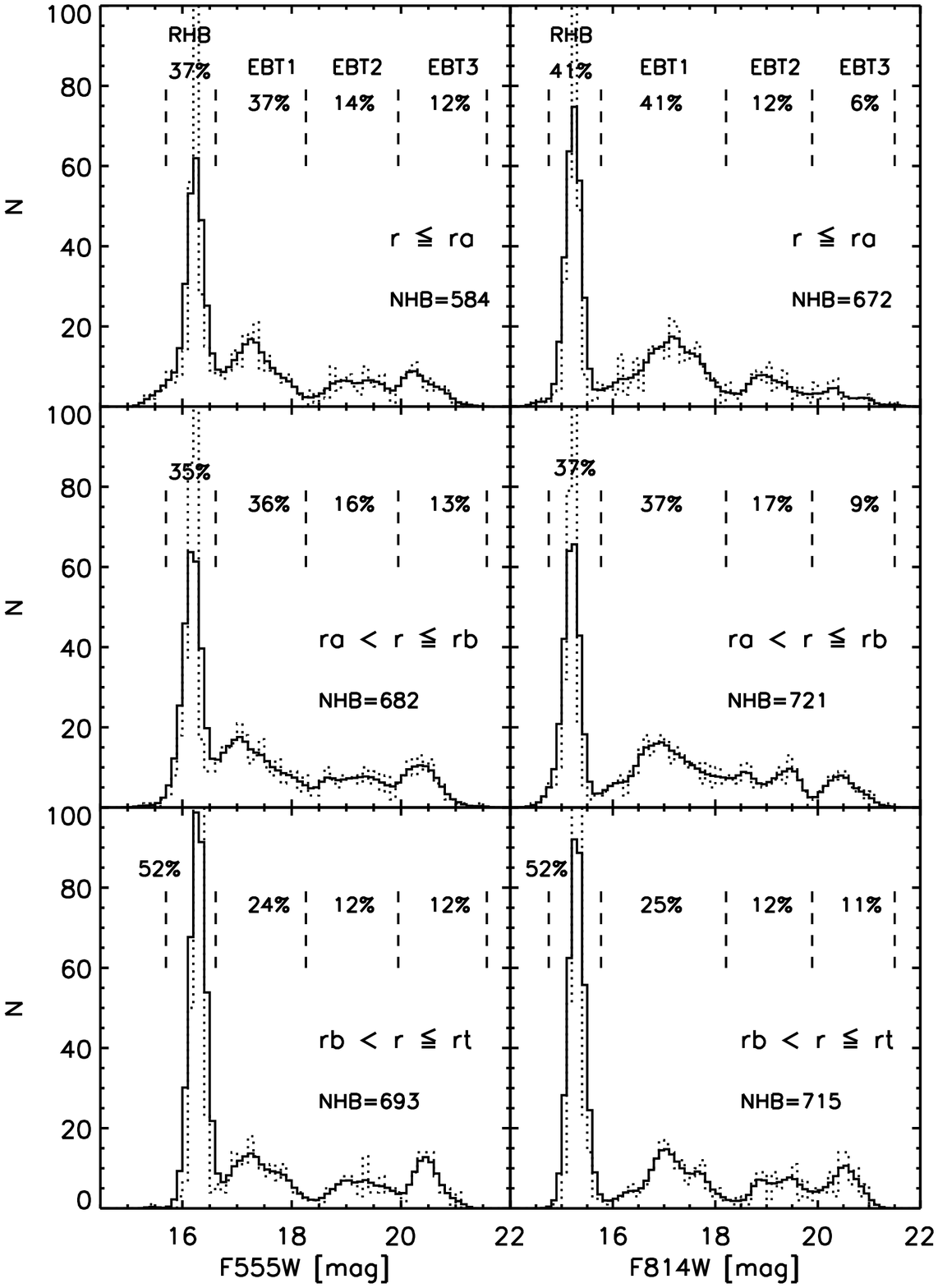}
\vspace*{0.60truecm}
\caption{Luminosity function of HB stars in the $F555W$ (left) and in 
the $F814W$-band (right). From top to bottom the three panels display 
the luminosity function in different radial bins. The dotted lines display 
the observed distributions, while the solid lines the smoothed luminosity 
function obtained using a Gaussian kernel. The vertical dashed lines show 
the magnitude ranges adopted to select the HB subgroups. The number 
of HB stars per bin and the relative fractions of HB subgroups are also 
labeled.   
}
\end{center}
\end{figure}

Data plotted in the pure optical CMDs---panel d) of Fig.~2 and panel a), b), c)
of Fig.~3---show that only a small fraction of the stars we have identified as
probable HB stars could be confused with either MS or RG stars.  Another small
fraction of the HB sample lies among the stars located between the HB and the
MS-RG fiducial cluster sequence.  These stars appear to be normal HB stars in
optical-UV CMDs.  They were labeled ``HB peculiar'' by Castellani et al.\ (2006)
and are probably either chance blends or physical binaries (Sandiquist \& Hess
2008).  

Overall, we ended up with a sample of $\approx 2,000$ HB stars distributed over
the entire body of the cluster. Note that the current sample is almost a factor
of two larger than the sample collected by Castellani et al.\ (2006). To
investigate their radial distribution, we split the total sample into three
subgroups, namely $r \le r_a=0.59$,  $r_a < r \le r_b=1.37$ and $r_b < r \le
r_t=15.5$ arcmin. The reason for specifically select these radial limits is twofold:
{\em i)} each radial bin includes approximately 1/3 of the entire sample; {\em
ii)} the two inner radial bins are based on HST data alone.

The lefthand panels of Fig.~4 show the Luminosity Function (LF) of the HB stars
in the three different radial bins. The observed distributions (dotted lines)
were smoothed using a conservative Gaussian kernel with standard deviation equal
to three times the formal photometric uncertainties of individual stars to
produce the solid curves. The vertical dashed lines show the range in apparent
magnitude for the different HB groups, namely the RHB, which dominates the
narrow peak, and the three EHB subgroups.  These data (see also Table~1)
indicate that, within the errors, the four HB subgroups follow the same radial
distribution when moving from the center to the outskirts of the cluster. 
Moreover, more than 50\% of all HB stars belong to the blue tail everywhere in
the cluster.  Data plotted in the right panels of Fig.~4 show that $F814W$-band
LFs show very similar trends for the four subgroups when moving from the center
to the outermost  cluster regions (see also Table~1). 

The apparent increase in the relative fraction of RHB stars in the external
radial bin is caused by field star contamination. To evaluate their contribution
on a quantitative basis we selected a region located well beyond the tidal
radius (r$\approx$19.4 arcmin), and using both the $F555W$,$F555W$-$F814W$ and
the $F555W$,$F336W$-$F814W$ CMD we found that we expect $\approx 52\pm7$ field
stars in the CMD box we adopted to define the RHB.  The uncertainty claimed here
accounts only for the Poisson uncertainty, and not for any (unexpected)
nonuniform distribution of field stars. Once we subtract these stars from the
external RHB group the relative fraction is still larger than for the inner two
zones, but the difference is now smaller than 1.5 $\sigma$.  The external EBT1
group may show a slight deficiency compared to the inner zones, but again
the anomaly is only of order $\sim1.5\sigma$ ($F555W$) to $\sim2\sigma$
($F814W$).  The other subgroups present similar differences, with no obvious
overall pattern.   

The above experiments indicate that the quantitative HB morphology in \ngc{2808}
does not show any pronounced radial trend. Moreover, the {\it qualitative\/} HB
morphology (i.e., clumps and gaps) also does not change when moving from the
very center to the outermost cluster regions, in agreement with the findings of
Walker (1999) and Bedin et al.\ (2000). This evidence indicates either that the
physical mechanism(s) driving the origin of the extended blue tail is at work
across the entire body of the cluster, or that any putative subpopulations
within the cluster are well mixed, in contrast with the predictions of D'Ercole
et al.\ (2008; see also Yoon et al.\ 2008). Moreover, the apparent constancy of
the HB with radius suggests that responsibility for the radial change in the
ratio of HB to RG stars (Castellani et al.\ 2006) may lie with the RGs. 
% Point A 
To validate this working hypothesis we performed detailed counts of bright
RGs using HST CMDs for the inner regions (see panel d of Fig~2 and panels a,b
of Fig.~3), and ground-based CMDs for the area located between the inner
regions and the tidal radius (panels c,d of Fig.~3). Following Castellani et al.\
(2006) we adopted $F555W_{\rm ZAHB}$=16.43 (VEGAMAG) and the same radial bins
adopted for HB star counts. The field star contamination in the external radial
bin was estimated using the same approach devised for RHB stars and we expect
to find $\approx 49\pm7$ field stars in the CMD box adopted to define bright
RGs. Data listed in Table~1 show that the number of RGs steadly decreases
from 444 for r$\le r_\alpha$ to 232 for  $r_\beta$$\le$ r $\le$$r_t$. The
total number of RGs we detected is 40\% larger than the number found by
Castellani et al.\ (2006). Interestingly enough, the $R$ parameter, i.e.\
the ratio between HB and RG stars brighter than the ZAHB, increases from
$1.32\pm0.08$ to $2.76\pm0.21$ (see Table~1). The difference is larger than
$5\sigma$ and confirms the trend found by Castellani et al.\ (2006). To further
constrain these findings we performed the same RG counts, but using the
$F814W$-band. Data listed in Table~1 show the same trend for RGs and the 
difference in the $R$ parameter between the innermost and the outermost
radial bin is larger than $4\sigma$. The current evidence indicates a 
deficiency of bright RGs in the outermost cluster regions.

\section{Acknowledgments}
It is a real pleasure to thank an anonymous referee for his/her positive 
comments and insights. We also thank S. Cassisi for many useful discussions
and a detailed reading of an early draft of this paper.
This project was partially supported by MIUR PRIN~2007 (PI: G. Piotto), 
INAF PRIN~2006 (PI: M. Bellazzini), Fondecyt Regular \#1085278, 
the Fondap Center for Astrophysics \#15010003, and by the BASAL CATA. 

%--------------------------------------------------------------------------------------

%%%%%%%%%%%%%%%%%%%%%%%%%%%%%%%%%%%%%%%%%%%%%%%%%%%%%%%%%%%%%%%%%%%%%%%%%%%%%%%%%
\begin{deluxetable}{l r r r r r r c}
\tablewidth{0pt}
\tabletypesize{\scriptsize}
\tablecaption{The HB and RGB star counts in different radial bins. 
Numbers in brackets are the relative fractions of the HB subgroups.}\label{}
\tablehead{
\colhead{Radius}   &
\colhead{N(RHB)}   &
\colhead{N(EBT1)}  &
\colhead{N(EBT2)}  &
\colhead{N(EBT3)}  &
\colhead{$N(HB)$}  & 
\colhead{$N(RGB)$\tablenotemark{a}} & 
\colhead{$R$\tablenotemark{b}}
}
\startdata
\multicolumn{8}{c}{Star counts based on the $F555W$-band LF}\\
$ r \le r_a $       & 217 ($37\pm3$\%) & 217 ($37\pm3$\%) & 81 ($14\pm2$\%)  & 69 ($12\pm1$\%) & 584  & 444  & $1.32\pm0.08$ \\

$ r_a < r \le r_b $ & 238 ($35\pm3$\%) & 244 ($36\pm3$\%) & 112 ($16\pm2$\%) & 88 ($13\pm1$\%) & 682  & 340  & $2.01\pm0.13$ \\

$ r_b < r \le r_t $ & 360 ($52\pm3$\%) & 164 ($24\pm2$\%) & 84 ($12\pm1$\%)  & 85 ($12\pm1$\%) & 693  & 281  & $2.47\pm0.17$ \\

Total               & 815 ($42\pm2$\%) & 625 ($32\pm2$\%) & 277 ($14\pm1$\%) & 242($12\pm1$\%) & 1959 & 1065 & $1.84\pm0.07$ \\

$\it r_b < r \le r_t$\tablenotemark{c}   & \it 308 ($\it 48\pm4$\%) & \it 164 ($\it 26\pm2$\%) & \it 84 ($\it 13\pm1$\%)  & \it 85 ($\it 13\pm1$\%) & \it 641  & \it 232 & $\it 2.76\pm0.21$ \\

\it Total\tablenotemark{c}              & \it 763 ($\it 40\pm3$\%) & \it 625 ($\it 33\pm2$\%) & \it 277 ($\it 14\pm1$\%) & \it 242 ($\it 13\pm1$\%) & \it 1907 & \it 1016 & $\it 1.88\pm0.07$ \\

\multicolumn{8}{c}{Star counts based on $F814W$-band LF\tablenotemark{d}}\\

$ r \le r_a $       & 273 ($41\pm3$\%) & 273 ($41\pm3$\%) & 86 ($12\pm2$\%)  & 40 ($6\pm1$\%) & 672  & 439  & $1.53\pm0.09$ \\

$ r_a < r \le r_b $ & 262 ($37\pm3$\%) & 269 ($37\pm3$\%) & 124 ($17\pm2$\%) & 66 ($9\pm1$\%) & 721  & 344  & $2.10\pm0.14$ \\

$ r_b < r \le r_t $ & 368 ($52\pm3$\%) & 180 ($25\pm2$\%) & 85 ($12\pm1$\%)  & 82 ($11\pm1$\%) & 715 & 298  & $2.40\pm0.17$ \\

Total               & 903 ($43\pm2$\%) & 722 ($34\pm2$\%) & 295 ($14\pm1$\%) & 188 ($9\pm1$\%) & 2108 & 1081  & $1.95\pm0.07$ \\

$ r_b < r \le r_t $\tablenotemark{c} & \it 316 ($48\pm4$\%) & \it 180 ($27\pm2$\%) & \it 85 ($13\pm1$\%)  & \it 82 ($12\pm1$\%) & \it 663 & \it 249 & $\it 2.66\pm0.20$ \\

\it Total\tablenotemark{c}           & \it 851 ($42\pm3$\%) & \it 722 ($35\pm2$\%) & \it 295 ($14\pm1$\%)  & \it 188 ($9\pm1$\%) & \it 2056 & \it 1032 & $\it 1.99\pm0.08$ \\
\enddata
\tablenotetext{a}{Star counts of RGs brighter than the ZAHB, i.e.\ RGs brighter than 
either $F555W$=16.43 or $F814W$=15.33 mag.}
\tablenotetext{b}{R parameter, i.e.\ the ratio between HB and RGs brighter than the ZAHB. 
The error only accounts for the Poisson uncertainty.}
\tablenotetext{c}{Star counts of the external radial bin cleaned for field star contamination in the RHB group.}
\tablenotetext{d}{The limits adopted in the $F814W$-band to identify the HB subgroups are: RHB -- redder than RR Lyrae and $14.75\le$$F814W$$\le15.77$, EBT1 -- bluer than RR Lyrae and 
$15.77<$$F814W$$\le18.21$, EBT2 -- $18.21<$$F814W$$\le19.89$, 
EBT3 -- $19.89<$$F814W$$\le21.50$.}
\end{deluxetable}

\end{document}